\documentclass[10pt]{article}
\usepackage{graphicx}
\usepackage{placeins}
\usepackage{hyperref}
 \usepackage{lineno}

\def\Title#1{\begin{center} {\Large #1 } \end{center}}
\def\Author#1{\begin{center}{ \sc #1} \end{center}}
\def\Address#1{\begin{center}{ \it #1} \end{center}}

\newcommand\pubblock{\rightline{\begin{tabular}{l} Proceedings of the Fifth Annual LHCP\\ \pubnumber\\
         \pubdate  \end{tabular}}}

\newenvironment{Abstract}{\begin{quotation} \begin{center} 
             \large ABSTRACT \end{center}\bigskip 
      \begin{center}\begin{large}}{\end{large}\end{center} \end{quotation}}

\newenvironment{Presented}{\begin{quotation} \begin{center} 
             PRESENTED AT\end{center}\bigskip 
      \begin{center}\begin{large}}{\end{large}\end{center} \end{quotation}}

\def\beq{\begin{equation}}
\def\eeq#1{\label{#1}\end{equation}}
\def\eeqn{\end{equation}}

\def\beqa{\begin{eqnarray}}
\def\eeqa#1{\label{#1}\end{eqnarray}}
\def\eeqan{\end{eqnarray}}

\let\bar=\overbar

\def\Dslash{\not{\hbox{\kern-4pt $D$}}}
\def\dslash{\not{\hbox{\kern-2pt $\del$}}}

\def\msb{{\bar{\ssstyle M \kern -1pt S}}}

\usepackage{color}

\newcommand\pubnumber{ ATL-PHYS-PROC-2017-112 }
\newcommand\pubdate{\today}

\def\affiliation{
On behalf of the ATLAS Collaboration, \\
Department of Physics \\
University of Washington, Seattle}

\begin{document}

%\linenumbers

% large size for the first page
\large
\begin{titlepage}
\pubblock

%% Change the title, name, abstract
%% Title 
\vfill
\Title{  Searches for Dark Matter in ATLAS  }
\vfill

%  if you need to add the support use this, fill the \support definition above. 
%   \Author{ FIRSTNAME LASTNAME \support }
\Author{ CRISTIANO  ALPIGIANI}
\Address{\affiliation}
\vfill
\begin{Abstract}

Although the existence of dark matter is well established by many astronomical measurements, its nature still remains one of the unsolved puzzles of particles physics. The unprecedented energy reached by the Large Hadron Collider (LHC) at CERN has allowed exploration of previously unaccessible kinematic regimes in the search for new phenomena. An overview of most recent searches for dark matter with the ATLAS detector at LHC is presented and the interpretation of the results in terms of effective field theory and simplified models is discussed. The exclusion limits set by the ATLAS searches are compared to the constraints from direct dark matter detection experiments.

\end{Abstract}
\vfill

% DO NOT CHANGE 
\begin{Presented}
The Fifth Annual Conference\\
 on Large Hadron Collider Physics \\
Shanghai Jiao Tong University, Shanghai, China\\ 
May 15-20, 2017
\end{Presented}
\vfill
\end{titlepage}
\def\thefootnote{\fnsymbol{footnote}}
\setcounter{footnote}{0}
%

% normal size for the rest
\normalsize 

%% Your paper should be entered below. 

\section{Introduction}

The nature of Dark Matter~(DM) still remains one of the largest open questions in particle physics. There is very strong astrophysical evidence for it coming especially from the observation of its gravitational effects at large distance scales \cite{ref:Planck}. All of these experiments have measured the universal DM density to be approximately five times that of normal matter, but there is still no experimentally confirmed theory of its origin. The DM problem calls for new physics beyond the Standard Model (SM) and the most prominent theories assume that the DM is a type of Weakly Interacting Massive Particle (WIMP), meaning that it weakly couples to the SM, as motivated by the so-called WIMP miracle where such a particle naturally reproduces the observed DM relic density. If DM is a WIMP, then the weak coupling between the SM and DM particles should be observable in the experiments like ATLAS. 

The DM searches can be classified in three categories: \emph{direct detection} performed measuring the DM recoiling off of SM particles, \emph{indirect detection} where experiments look for SM by-products of galactic DM-DM annihilation, and finally \emph{collider production} of DM through collisions of particles in the accelerators.

ATLAS searches for the direct production of DM focus on essentially two main strategies. The first one searches for events in which DM is pair-produced in association with another object; this topology is defined as $E_\mathrm{T}^\mathrm{miss}+X$, where the $X$ represents object(s) that interact with the detector. These objects are typically from Initial State Radiation (ISR) sources, although in some cases they may come from a mixing with the mediator. This ISR object is necessary, as the production of DM without other particle(s) is invisible to the detector, while visible object(s) plus DM will appear as a significant imbalance in the transverse energy ($E_\mathrm{T}^\mathrm{miss}$) of the event. The second strategy looks for a mediator, which could connect the SM to the DM. An example is the search for resonances in the $m_{jj}$ spectrum of di-jets events, where it is assumed that the mediator couples to light quarks or gluons, and an observed resonance could be interpreted as the production of a new dark mediator.

With the data collected in 2015 and 2016, no statistically significant excess beyond SM expectations is observed in any of the topologies discussed in the next sections, and the result are interpreted in terms of Effective Field Theories (EFT) and simplified models, connecting the two SM inputs to the two DM outputs via a single vector, axial-vector, or scalar mediator.

\begin{figure}[!h]
\centering
\includegraphics[width=0.7\textwidth]{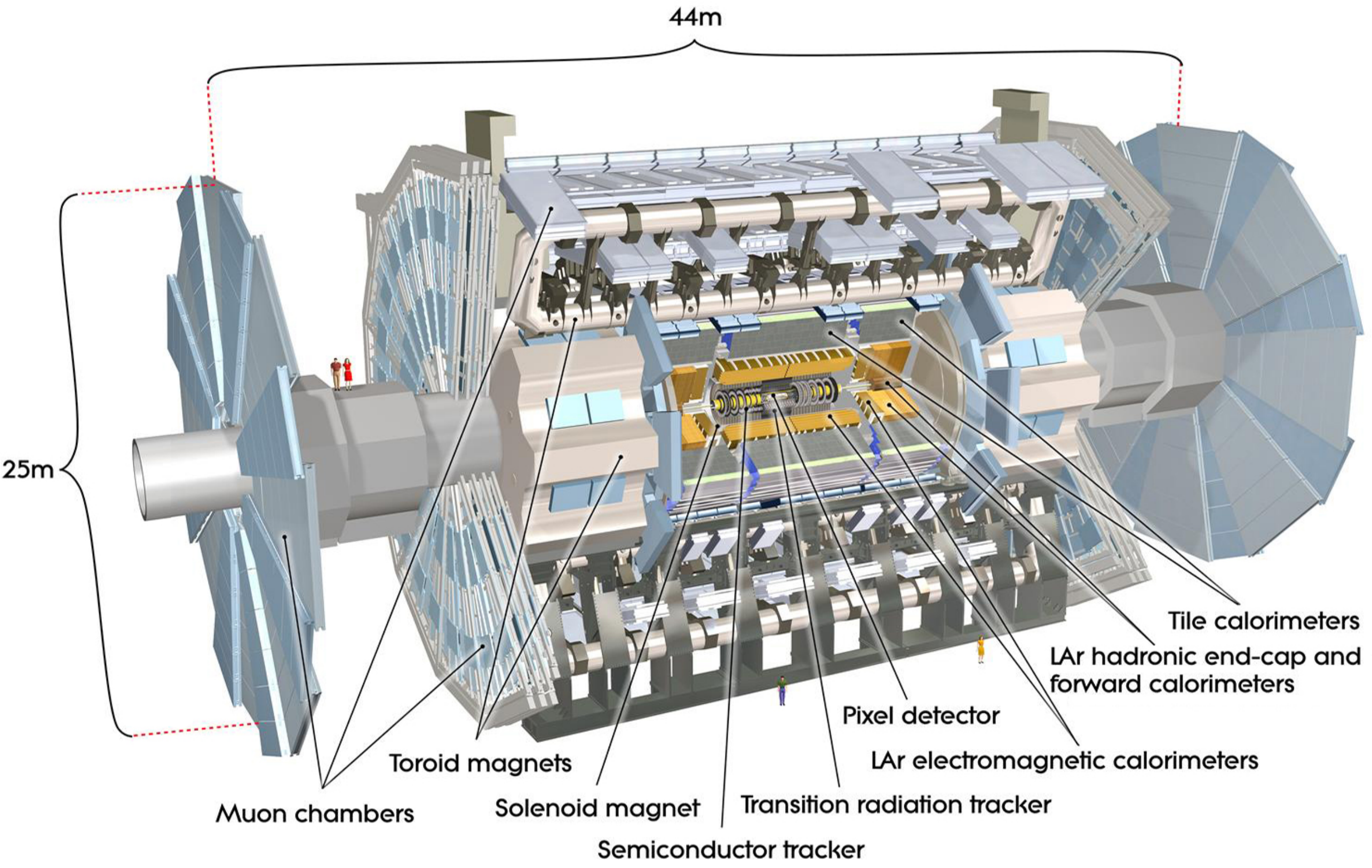}
\vspace{-0.2 cm}
\caption{Overview of the ATLAS detector (taken from Ref.~\cite{ref:atlas}).}
\label{fig:atlas}
\end{figure}

\section{The ATLAS experiment}

The ATLAS (A Toroidal Lhc ApparatuS) detector~\cite{ref:atlas,ref:ATLASdetector} is a general purpose experiment operating at the Large Hadron Collider (LHC) at CERN, and it is divided in three main components (see Figure \ref{fig:atlas}). Moving outward from the interaction point, the detector consists of a tracking system (inner detector) able to measure the directions and momenta of the charged particles. The calorimeter system measures the energies of the electrons, photons and hadrons. The muon spectrometer measures the momentum and position of the muons that have enough energy to reach it. Furthermore, a magnet system provides a magnetic field, allowing the tracker and the muon spectrometer to perform measurements of the momentum of charged particles.

\section{Dark matter with mono\,-photon}

The mono-photon analysis \cite{ref:monoPhoton} uses data corresponding to an integrated luminosity of 36.1 fb$^{-1}$ at 13 TeV. It is performed looking for events with a single energetic photon, large $E_\mathrm{T}^\mathrm{miss}$, no leptons and not more than one jet in the final state.

The background is dominated by electroweak production, $\gamma+Z(Z\rightarrow\nu\nu)$ where the photon comes from ISR. Secondary contributions come from $W(\rightarrow \ell\nu)\gamma$ and $Z(\rightarrow \ell\ell)\gamma$ production with unidentified electrons, muons or with $\tau \rightarrow \textrm{hadrons}+\nu_\tau$ decays and from  $\gamma+\textrm{jets}$ events. Contributions from $t\bar{t}+\gamma$ and events where a lepton or a jet is misidentified as a photon are negligible. 

Five different signal regions (SRs) are defined with different $E_\mathrm{T}^\mathrm{miss}$ requirements and a simultaneous fit in background-enriched control regions (CR) is performed to obtain the normalisation factors for the $W\gamma$, $Z\gamma$ and $\gamma$+jets backgrounds, which are then used to estimate backgrounds in the SRs.

\begin{figure}[!h]
\centering
\includegraphics[width=0.495\textwidth]{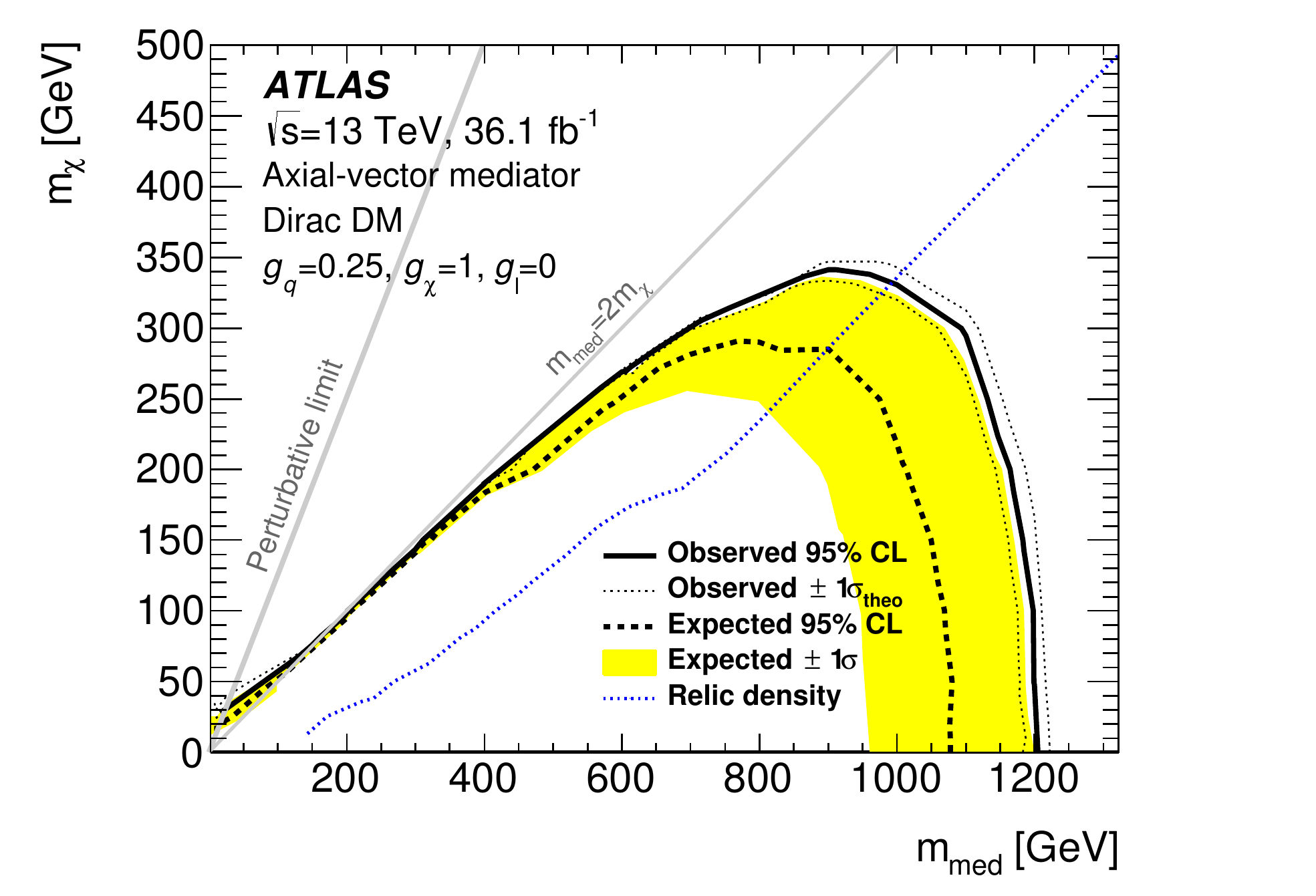}
\includegraphics[width=0.495\textwidth]{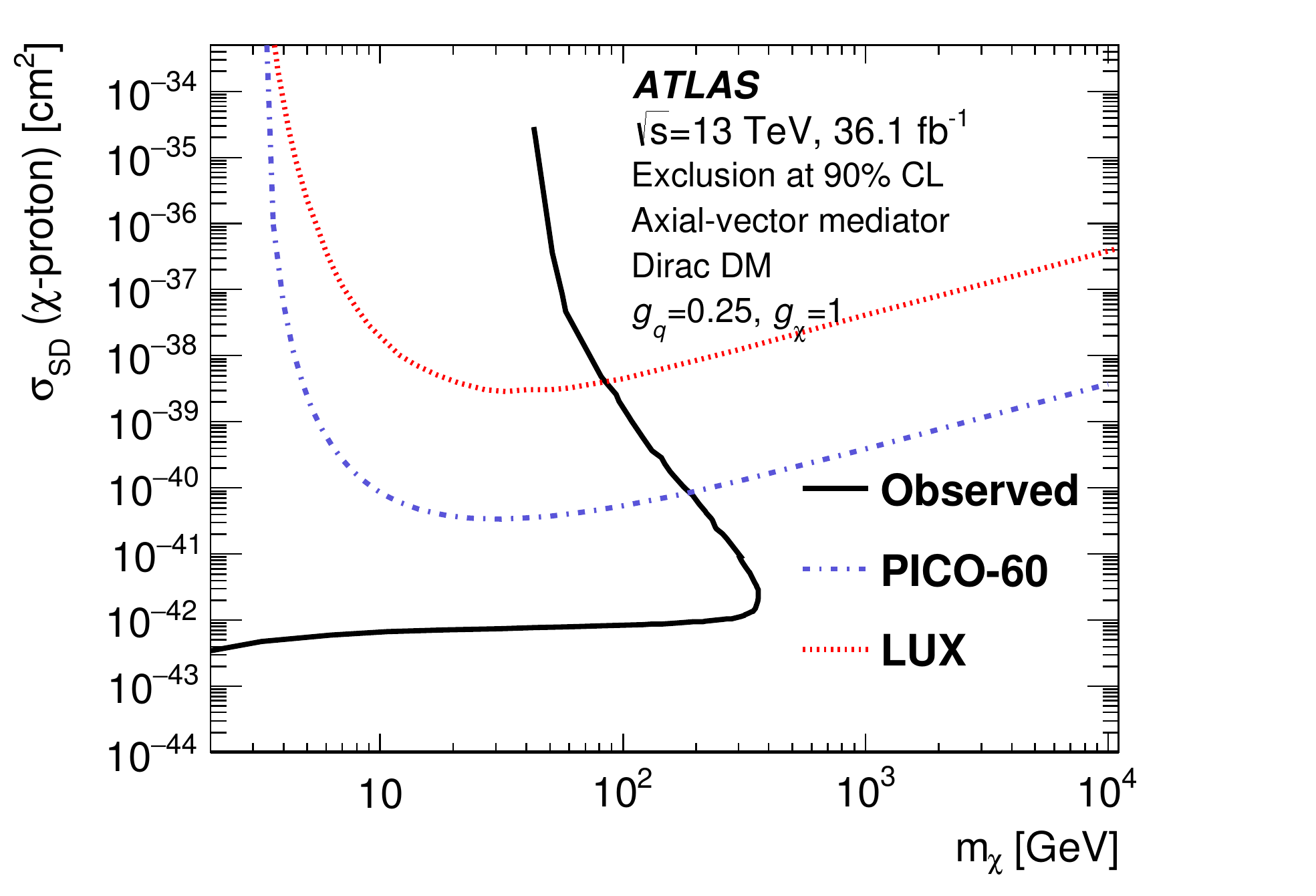}
\vspace{-0.8 cm}
\caption{Left: exclusion limits at 95\% CL for a simplified model of DM production with an axial-vector mediator, Dirac DM particles and couplings $g_q = 0.25$, $g_\chi = 1$ and $g_l=0$ as a function of the DM mass and the mediator mass (taken from Ref.~\cite{ref:monoPhoton}). Right: exclusion limit at 90\% CL on the $\chi\textrm{\,-\,proton}$ scattering cross section as a function of the DM mass $m_\chi$ in a simplified model of DM production involving an axial-vector operator, Dirac DM particles and couplings $g_q=0.25$ and $g_\chi=1$ (taken from Ref.~\cite{ref:monoPhoton}). The results obtained with the direct detection experiments Pico-60 \cite{ref:Pico60} and LUX \cite{ref:LUX} are shown for comparison.}
\label{fig:monoPhoton}
\end{figure}

The results of the search for the mono-photon\,+\,$E_\mathrm{T}^\mathrm{miss}$ are interpreted both in terms of EFT and a simplified model with an axial-vector mediator. Figure \ref{fig:monoPhoton} (left) shows the exclusion limits for a simplified model of DM production with an axial-vector mediator, Dirac DM particles, and couplings $g_q = 0.25$ and $g_\chi = 1$ as a function of the DM mass and the mediator mass. Figure \ref{fig:monoPhoton} (right) shows the exclusion limit on the $\chi\textrm{\,-\,proton}$ scattering cross section as a function of the DM mass $m_\chi$ in a simplified model of DM production involving an axial-vector operator, Dirac DM particles and couplings $g_q=0.25$ and $g_\chi=1$. For the simplified dark-matter model considered, the current search~\cite{ref:monoPhoton} excludes axial-vector and vector mediators with masses below 750-1200 GeV for $\chi$ masses below 230-480 GeV at 95\% CL, depending on the couplings.

\section{Dark matter with mono\,-jets}

DM searches in the mono-jet channel have been performed in ATLAS using data at $\sqrt{s}=13$ TeV corresponding to an integrated luminosity of 36.1 fb$^{-1}$ collected in 2015 and 2016 \cite{ref:monoJet}.
The analysis selects events with a high $p_\mathrm{T}$ jet and large missing transverse energy $E_\mathrm{T}^\mathrm{miss}$ in the final state. Several inclusive and exclusive signal regions (SRs) with increasing requirements on the missing transverse energy $E_\mathrm{T}^\mathrm{miss}$ between 250 and 700 GeV are optimised. The expected background is dominated by $Z(\rightarrow \nu\nu)$+jets and $W$+jets production processes. The contributions of the background processes to the SRs are estimated using dedicated control regions.

\begin{figure}[h!]
\centering
\includegraphics[width=0.44\textwidth]{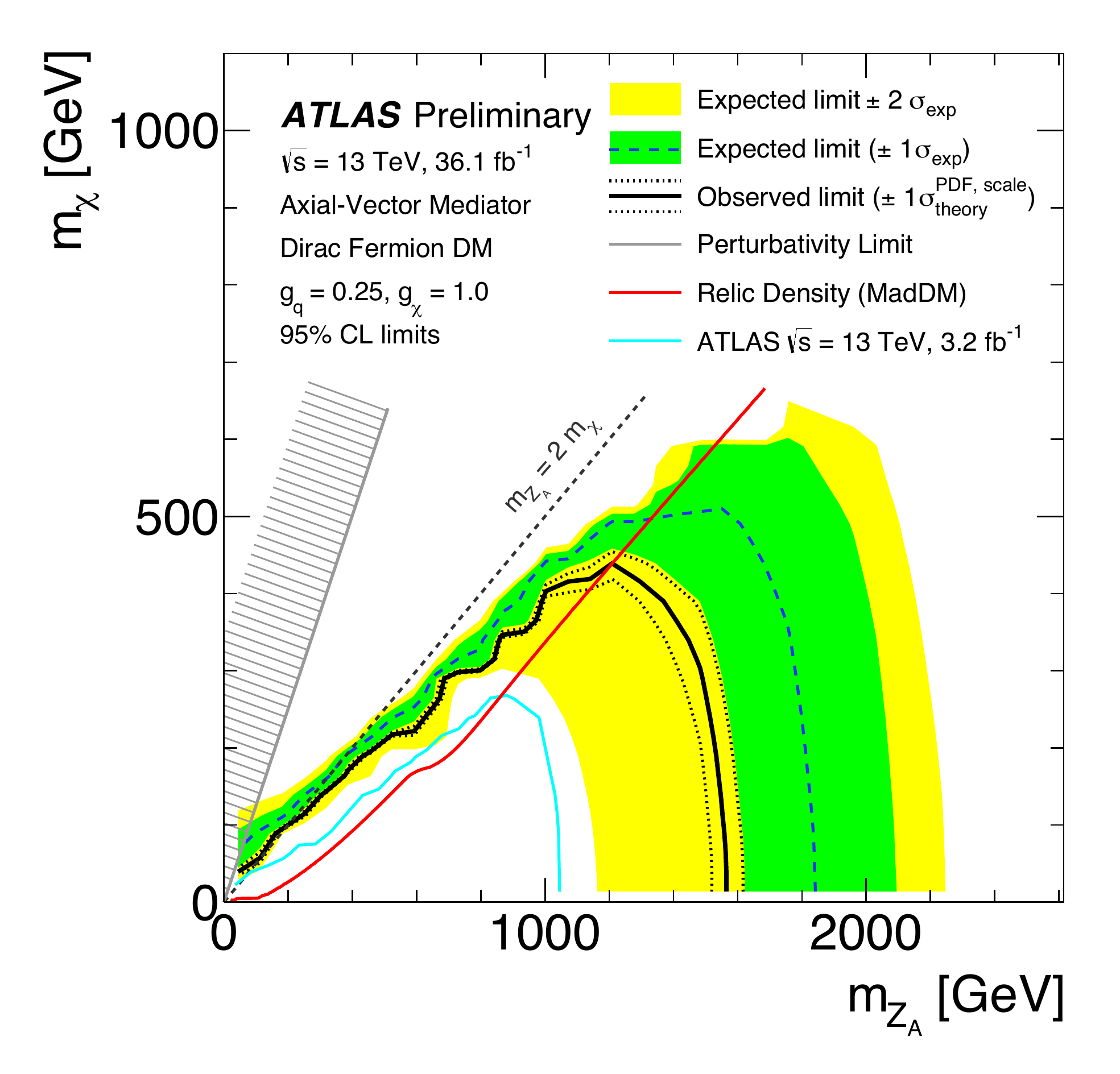}
\includegraphics[width=0.44\textwidth]{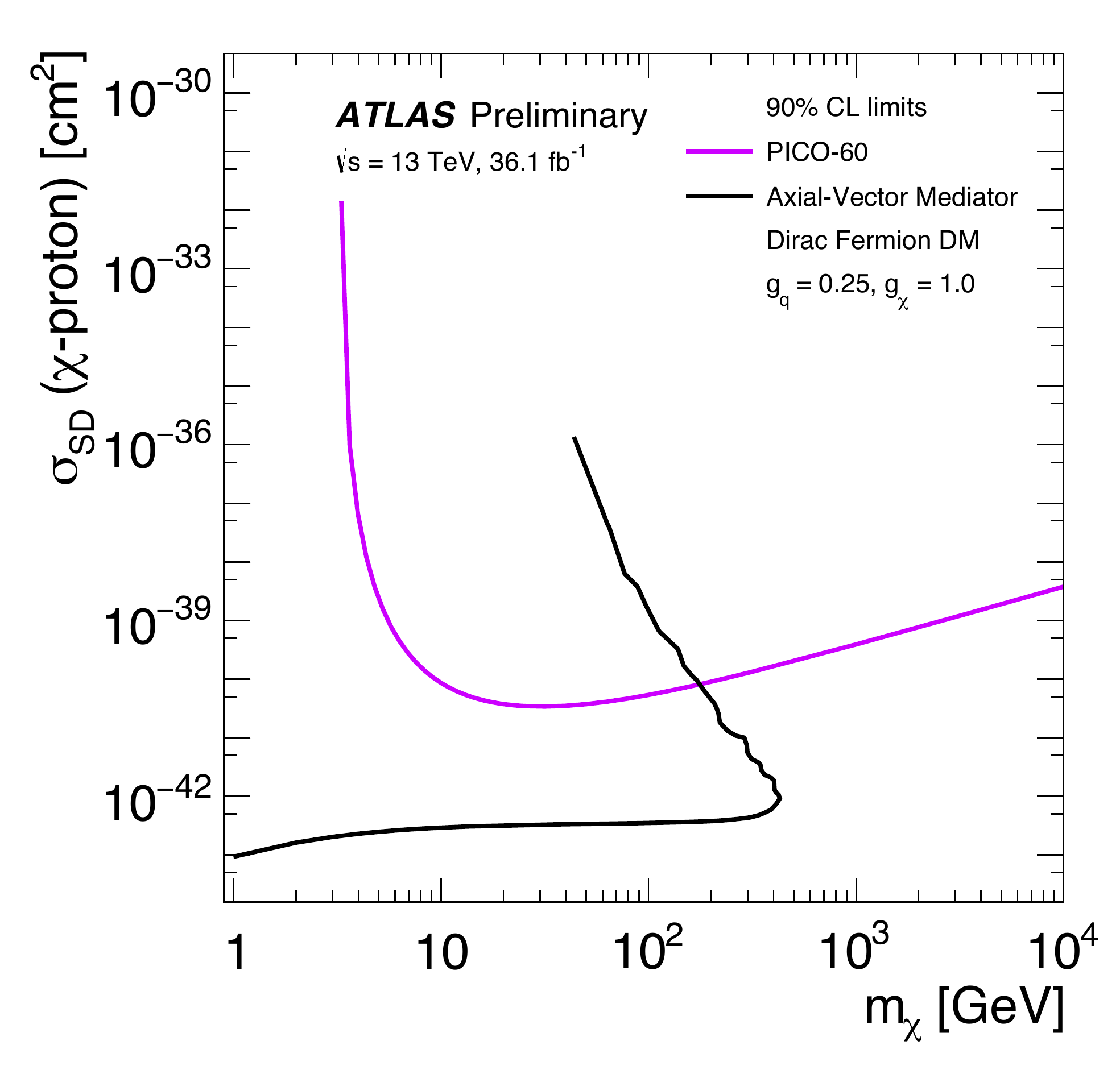}
\vspace{-0.5 cm}
\caption{Left: exclusion limits in the $m_\chi \textrm{-} m_{Z_A}$ parameter plane for a simplified model with an axial-vector mediator $Z_A$, Dirac DM particles, and couplings $g_q = 0.25$ and $g_\chi = 1$ (taken from Ref.~\cite{ref:monoJet}). Right: limits on the spin-dependent $\chi\textrm{-}\textrm{proton}$ scattering cross section in the context of the $Z^\prime$-like simplified model with axial-vector couplings ($g_q = 0.25$ and $g_\chi = 1$) as a function of $m_\chi$, compared to the constraints from direct DM detection PICO-60 experiment \cite{ref:Pico60} (taken from Ref.~\cite{ref:monoJet}).}
\label{fig:monoJet}
\end{figure}

No significant excess was found in data, and the results are translated into exclusion limits in models with large extra spatial dimensions, pair production of weakly interacting dark-matter candidates, and the production of supersymmetric particles in several compressed scenarios. Figure \ref{fig:monoJet} (left) shows observed limits in the $m_\chi \textrm{-} m_{Z_A}$ parameter plane for a simplified model with an axial-vector mediator $Z_A$, Dirac DM particles, and couplings $g_q = 0.25$ and $g_\chi = 1$. The models with mediator masses up to 1.55 TeV are excluded for $m_\chi=1$ GeV in the on-shell regime. Figure \ref{fig:monoJet} (right) reports the limit on the spin-dependent $\chi\textrm{-}\textrm{proton}$ scattering cross section as a function of $m_\chi$. The comparison is model-dependent and solely valid in the context of the particular $Z^\prime$-like model. In this case, stringent limits on the scattering cross section of the order of $10^{-43}$ cm$^2$ for WIMP masses below 10 GeV are inferred from this figure and complement the results from direct-detection experiments for $m_\chi<10$ GeV.

%% mono-V
%
\section{Dark matter with mono\,-$V$}

A search for DM using mono-jet is performed looking for a boosted boson that recoils against pair-produced DM particles. The analysis uses proton-proton collision data corresponding to an integrated luminosity of 3.2 fb$^{-1}$ at $\sqrt{s}=13$ TeV collected in 2015 \cite{ref:monoV}. To reconstruct the $W/Z$ boson in the boosted regime, the hadronic products of the produced quarks are captured by a high-$p_\mathrm{T}$ large-R jet. The expected background is dominated by $Z(\rightarrow\nu\nu)$+jet and $W$+jets production, with $W\rightarrow\tau\nu$ being the largest $W$+jets background. The kinematic distributions of these backgrounds are estimated using simulated event samples, and the normalisation is determined using control regions where the dark-matter signal is expected to be negligible. 

\begin{figure}[h!]
\centering
\includegraphics[width=0.48\textwidth]{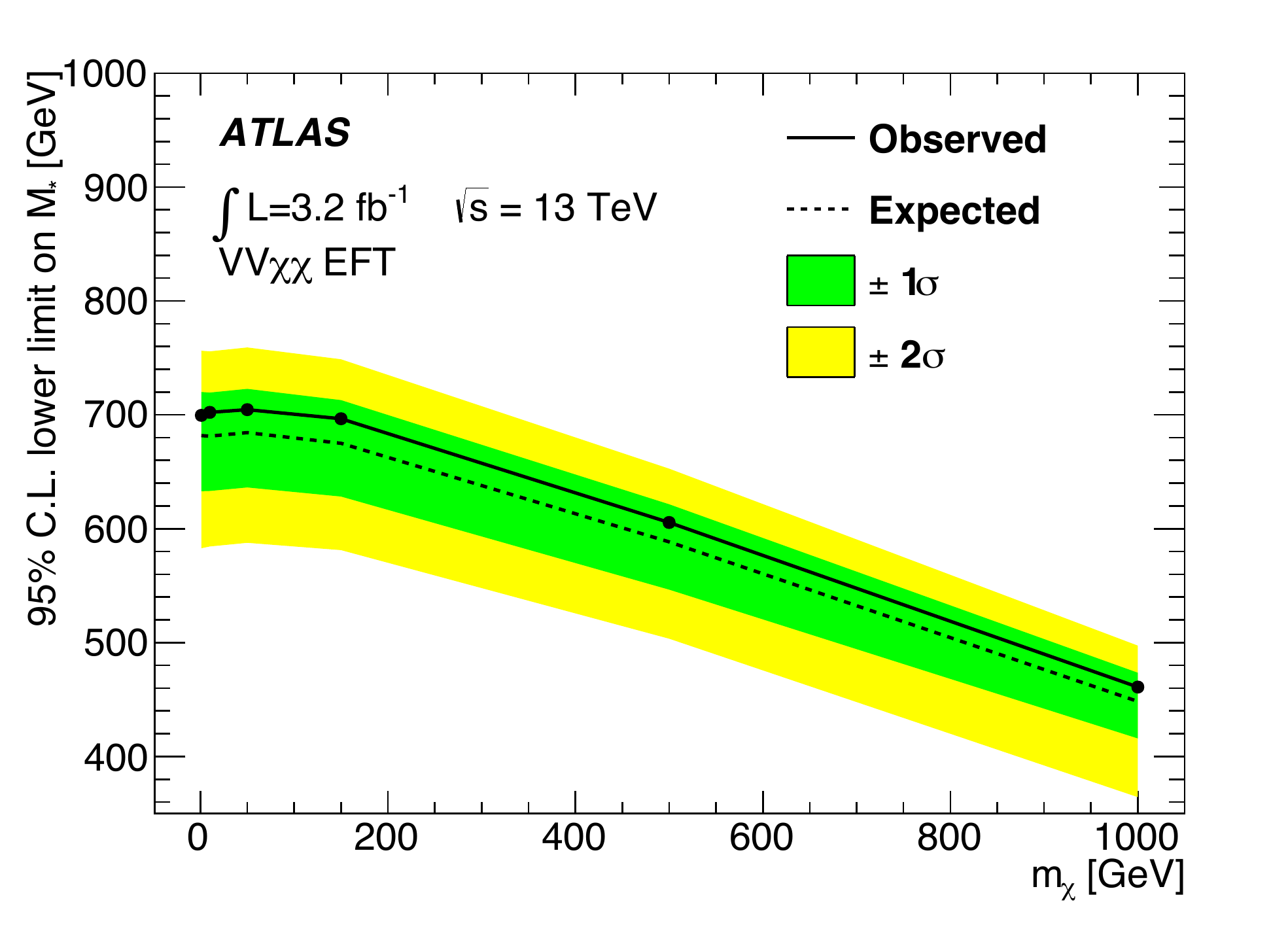}
\includegraphics[width=0.48\textwidth]{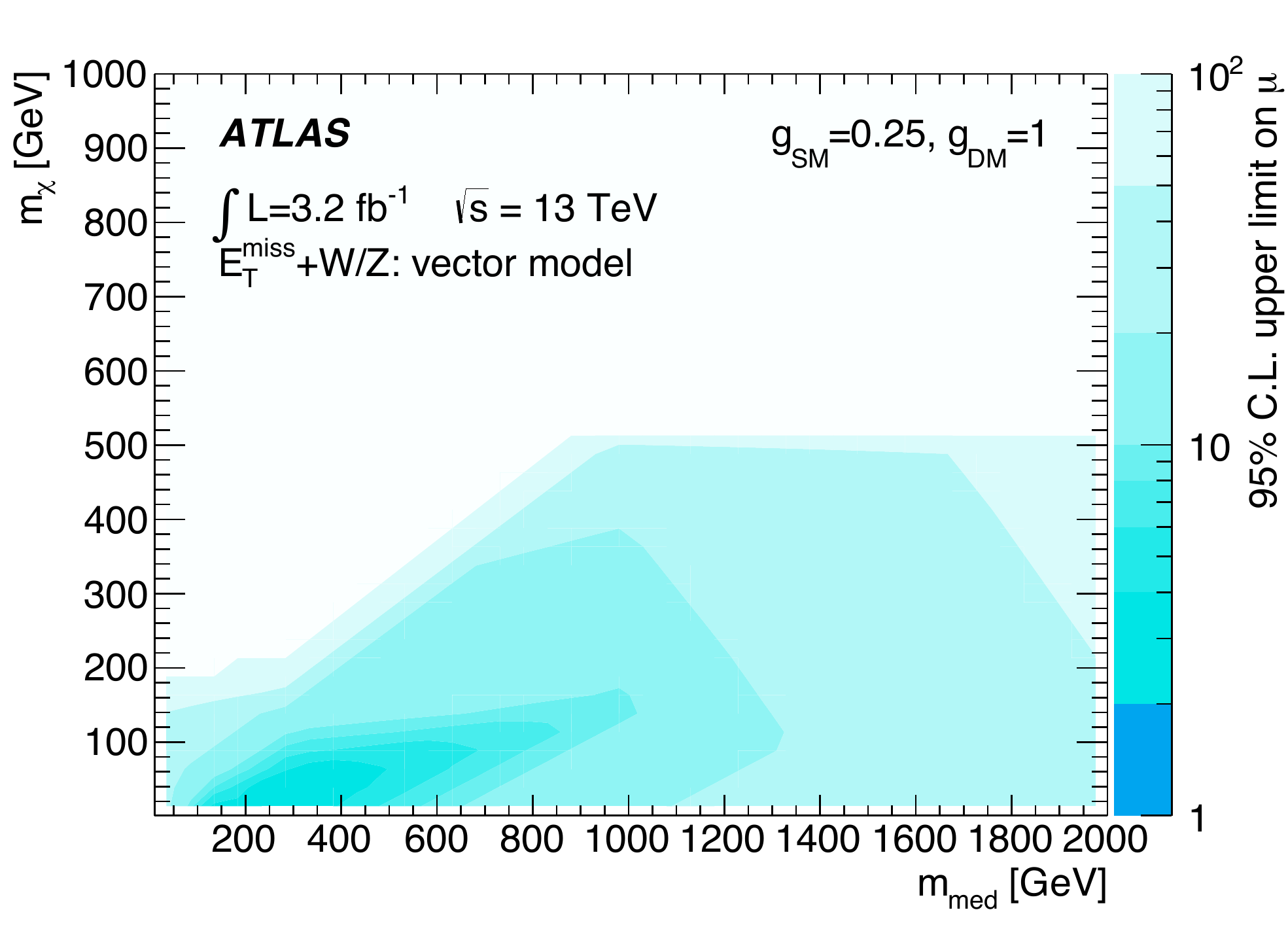}
\vspace{-0.5 cm}
\caption{Left: exclusion limits on the mass scale, $M_*$, of the $VV\chi\chi$ EFT model (taken from Ref.~\cite{ref:monoV}). Right: observed limit on the signal strength, $\mu$, of the vector-mediated simplified model (with couplings $g_\textrm{SM}=0.25$ and $g_\textrm{DM}=1$) in the plane of the dark-matter particle mass, $m_\chi$, and the mediator mass, $m_{\textrm{med}}$ (taken from Ref.~\cite{ref:monoV}). White areas indicate an upper limit at $\mu \ge 100$.}
\label{fig:monoV}
\vspace{-0.3 cm}
\end{figure}

No statistically significant excess over the Standard Model prediction is observed and limits are set both in term of EFT and simplified models. Figure \ref{fig:monoV} (left) shows the observed limit on the mass scale, $M_*$, of the $VV\chi\chi$ EFT model, while Figure \ref{fig:monoV} (right) reports the observed limit on the signal strength, $\mu$, of the vector-mediated simplified model in the plane of the dark-matter particle mass, $m_\chi$, and the mediator mass, $m_{\textrm{med}}$ (the model is generated with couplings $g_\textrm{SM}=0.25$ and $g_\textrm{DM}=1$).

\vspace{0.25 cm}

\begin{figure}[h!]
\centering
\includegraphics[width=0.438\textwidth]{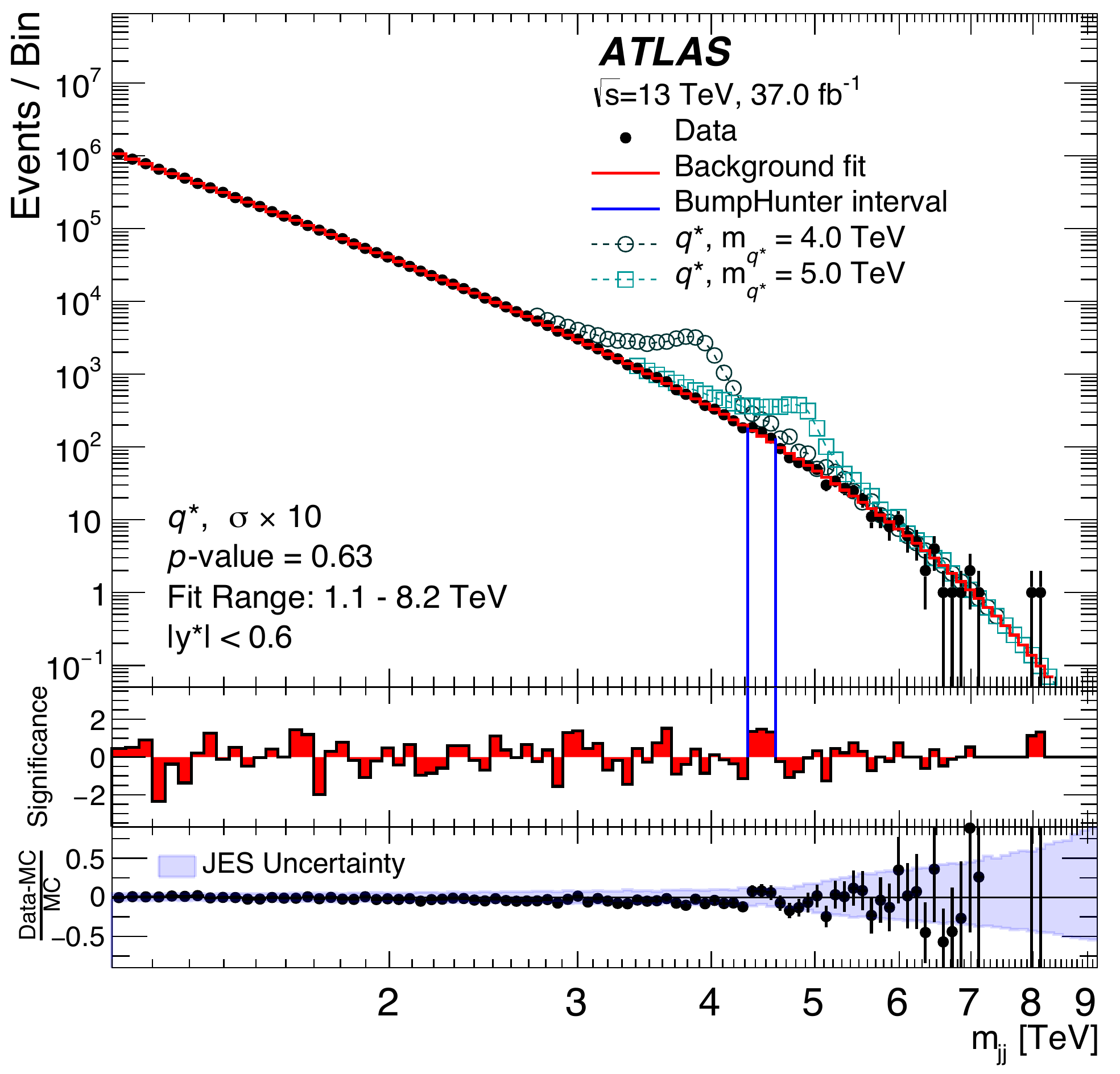}
\includegraphics[width=0.438\textwidth]{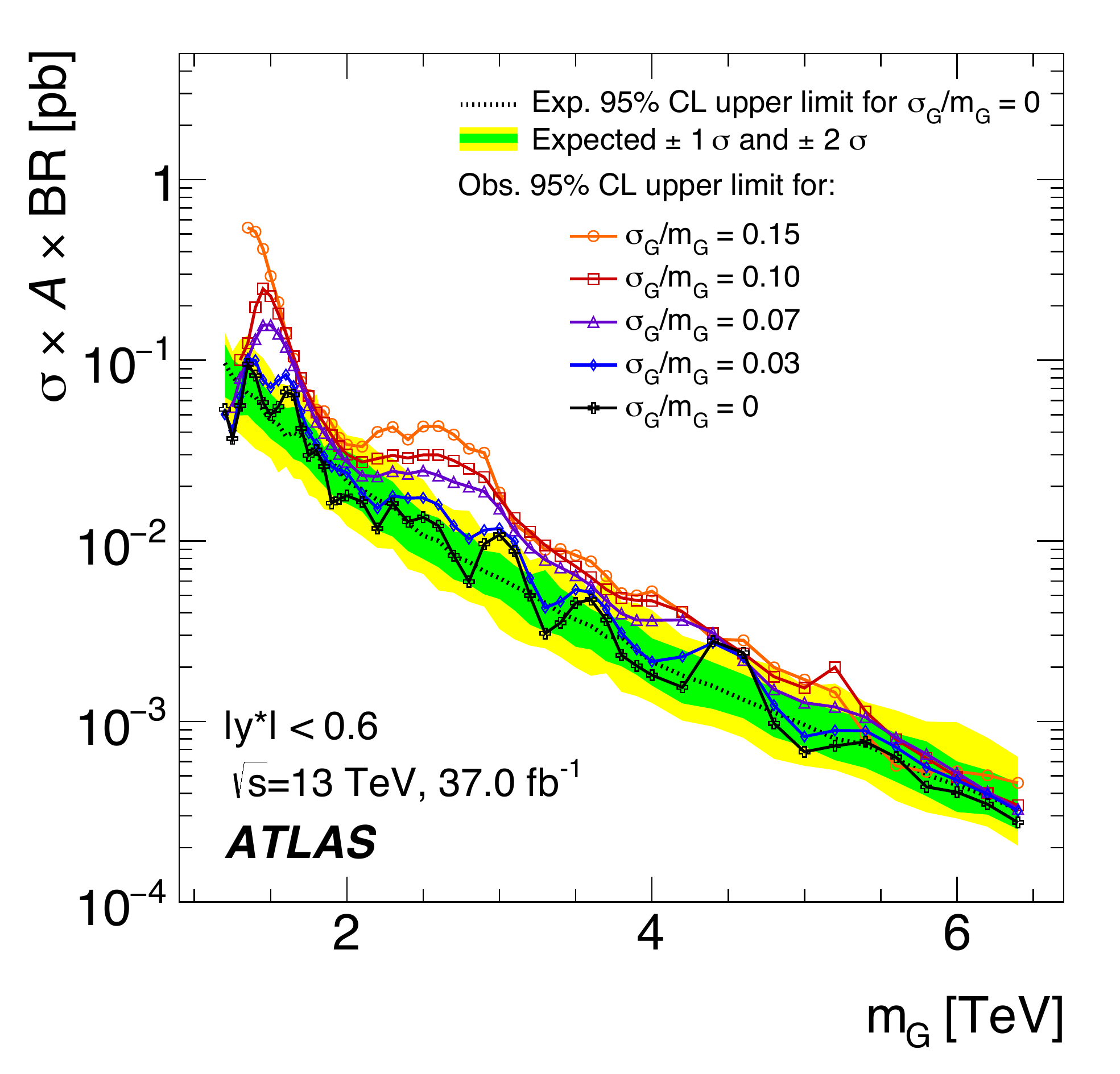}
\vspace{-0.4 cm}
\caption{Left: reconstructed di-jet mass distribution $m_{jj}$ (filled points) for events with $p_\mathrm{T} > 440\,(60)$ GeV for the leading (sub-leading) jet (taken from Ref.~\cite{ref:dijets}). 
%The spectrum with $|y^*| < 0.6$ is shown in for events above $m_{jj}$ = 1.1 TeV. 
The solid line depicts the background prediction from the sliding-window fit. The vertical lines indicate the most discrepant interval identified by the BumpHunter algorithm, for which the $p$-value is stated in the figure. The middle panel shows the bin-by-bin significances of the data-fit differences, considering only statistical uncertainties. The lower panel shows the relative differences between the data and the prediction of Pythia 8 simulation of QCD processes. The shaded band denotes the experimental uncertainty in the jet energy scale calibration. Right: upper limits obtained from the di-jet invariant mass distribution on cross-section times acceptance times branching ratio to two jets, $\sigma \times A \times BR$, for a hypothetical signal with a cross-section $\sigma_G$ that produces a Gaussian contribution to the particle-level $m_{jj}$ distribution, as a function of the mean of the Gaussian mass distribution $m_G$ (taken from Ref.~\cite{ref:dijets}). Observed limits are obtained for five different widths, from a narrow width to 15\% of $m_G$.}
\label{fig:dijets}
\end{figure}

\section{Dark matter with di-jets}

New phenomena beyond the SM can also be studied using di-jet events in proton-proton collision recorded at $\sqrt{s}=13$ TeV in 2015 and 2016, corresponding to integrated luminosities of 37 fb$^{-1}$. The analysis performed by the ATLAS experiment looks for resonances in the $m_{jj}$ spectrum and it is completely data-driven \cite{ref:dijets}.

The $m_{jj}$ distribution formed from the two leading jets in selected events is analysed looking for resonances from BSM phenomena. 
%The rapidity of an outgoing parton is defined as $y = 1/2 \cdot \ln[(E+p_z)/(E-p_z)]$, where E is its energy and $p_z$ is the component of its momentum along the z-axis. 
The rapidity difference $y^* = (y_1 - y_2)/2$, defined between the two leading jets, is invariant under Lorentz boosts along the $z$-axis and it is used to reduce the QCD background. The region $|y^*| < 0.6$ is used for the model-independent search phase, to set limits on generically-shaped signals, and to constrain the $q^*$, quantum black holes (QBH), $W^\prime$ and $Z^\prime$ benchmark models, all of whose distributions peak at $y^*=0$. The $W^*$ benchmark model, whose distribution peaks at $y^* > 1.0$, is constrained using a wider selection, $y^* < 1.2$, optimised for signals produced at more forward angles.

Figure \ref{fig:dijets} (left) shows the observed $m_{jj}$ distribution for events passing the $y*$ selection, overlaid with an example of the signal. The background estimate is illustrated by the solid red line and is derived from the sliding-window fitting method, improved with respect to the previous analysis. The background for the invariant mass spectrum is constructed bin-by-bin by performing a likelihood fit to the data in each window and using the fit value at the center of the window for the background description. No significant deviation is observed in data and limits, reported in Figure \ref{fig:dijets} (right), are set on the cross-section times acceptance times branching ratio to two jets for a hypothetical signal that produces a Gaussian contribution to the particle-level $m_{jj}$ distribution. This analysis excludes several types of signals as predicted by models of quantum black holes, excited quarks, and $W^\prime$, $W^*$ and $Z^\prime$ bosons. These results substantially extend the excluded ranges obtained using the 2015 dataset alone, with improvements ranging from 7\% for quantum black hole masses to 25\% for contact interaction scales to 40\% for $W^\prime$ boson masses.

\section{Summary}

\begin{figure}[h]
\centering
\includegraphics[width=0.68\textwidth]{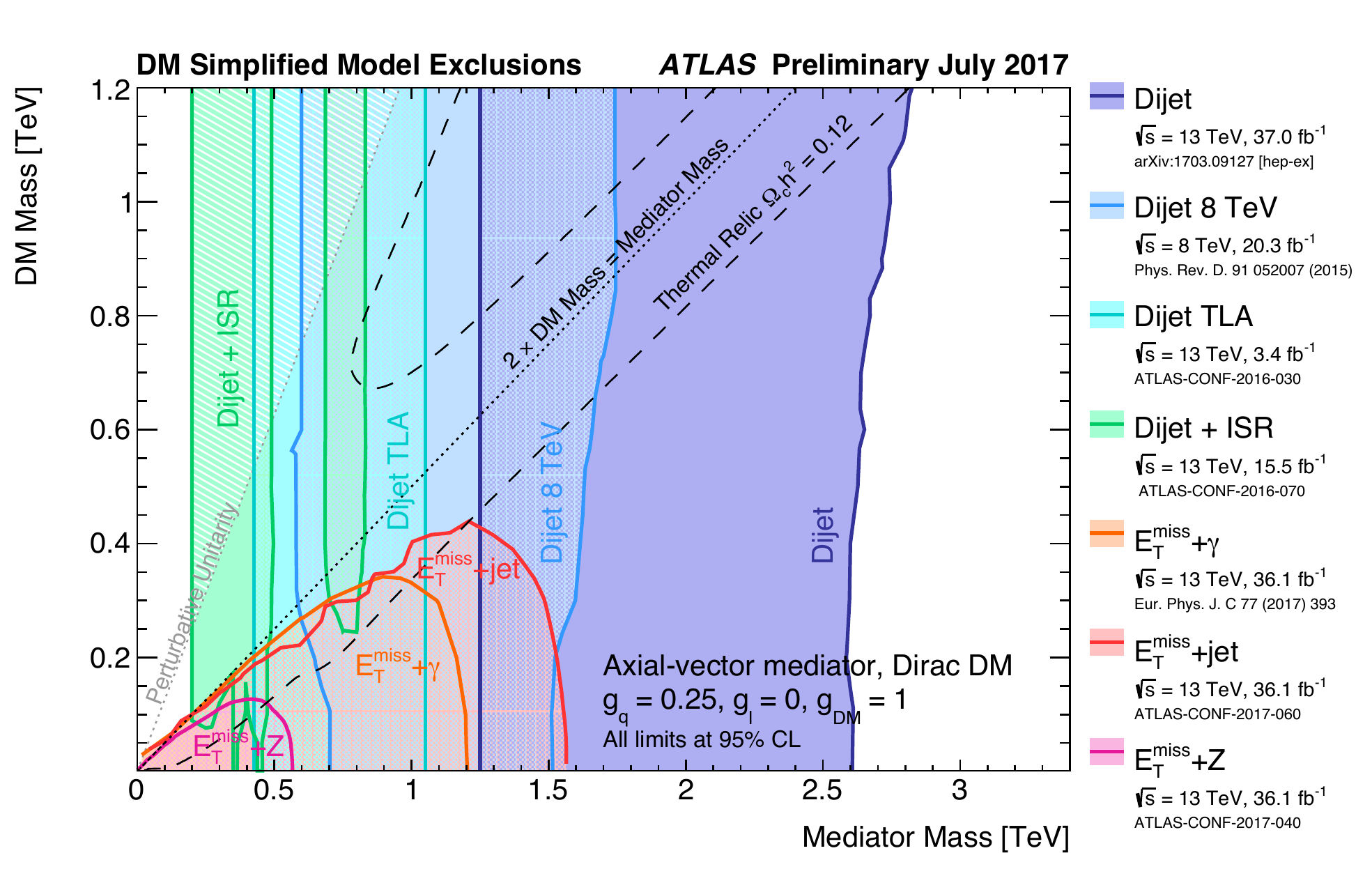}
\vspace{-0.5 cm}	
\caption{95\% CL excluded regions in a dark matter mass-mediator mass plane from a selection of ATLAS dark matter searches (taken from Ref.~\cite{ref:DMsummary}). Exclusions are computed for a dark matter coupling $g_\textrm{DM} = 1.0$ and a quark coupling $g_q = 0.25$ universal to all flavours. The lepton coupling $g_l$ in this model is set to zero. The results use 13 TeV data except for Phys. Rev. D91 052007 (2015), which was 8 TeV data. Dashed curves labeled ``thermal relic'' indicate combinations of dark matter and mediator mass that are consistent with a dark matter density of $\Omega_c = 0.12 \, h^2$ and a standard thermal history. Between the two curves, annihilation processes described by the simplified model deplete $\Omega_c$ below $0.12 \, h^2$. A dotted curve indicates the kinematic threshold where the mediator can decay on-shell into dark matter.}
\label{fig:DMModelExclusions}
\end{figure}

The search for Dark Matter is a key part of the ATLAS physics program and several analyses have been performed using both Run 1 and Run 2 datasets. A excellent summary, even if not comprehensive, is provided by Figure \ref{fig:DMModelExclusions} that shows the regions in a dark matter mass-mediator mass plane excluded at 95\% CL by a selection of ATLAS dark matter searches that use 8 and 13 TeV data, for one possible interaction between the Standard Model and dark matter \cite{ref:DMsummary}.

No evidence of physics beyond the Standard Model is observed so far by ATLAS, and limits are set on vector, axial-vector, and scalar simplified models of DM production.

%\clearpage
\FloatBarrier


\begin{thebibliography}{99}

%%
%%  bibliographic items can be constructed using the LaTeX format in SPIRES:
%%    see    http://www.slac.stanford.edu/spires/hep/latex.html
%%  SPIRES will also supply the CITATION line information; please include it.
%%
\bibitem{ref:Planck} Planck Collaboration 2014 Astron. Astrophys. 571 A16

\bibitem{ref:atlas} ATLAS Collaboration, \emph{The ATLAS Experiment at the CERN Large Hadron Collider}, JINST \textbf{3} 2008 S08003.
   
\bibitem{ref:ATLASdetector} ATLAS Collaboration, \emph{The ATLAS Inner Detector commissioning and calibration}, Eur. Phys. J. \textbf{C70} (2010), 787 - 821, arXiv:1004.5293 [physics.ins-det].

\bibitem{ref:monoPhoton}  ATLAS Collaboration, \emph{Search for dark matter at $\sqrt{s}=13$ TeV in final states containing an energetic photon and large missing transverse momentum with the ATLAS detector}, Eur. Phys. J. C 77 (2017) 393, arXiv:1704.03848 [hep-ex].

\bibitem{ref:Pico60} PICO-60 Collaboration, C. Amole et al., \emph{Dark Matter Search Results from the PICO-60 C$_3$F$_8$ Bubble Chamber}, Phys. Rev. Lett. 118 (2017) 251301, arXiv: 1702.07666 [astro-ph.CO].

\bibitem{ref:LUX} LUX Collaboration, D. S. Akerib et al., \emph{Results on the Spin-Dependent Scattering of Weakly Interacting Massive Particles on Nucleons from the Run 3 Data of the LUX Experiment}, Phys. Rev. Lett. 116 (2016) 161302, arXiv: 1602.03489 [hep-ex].

%\bibitem{ref:monoJet} ATLAS Collaboration, \emph{Search for new phenomena in final states with an energetic jet and large missing transverse momentum in pp collisions at $sqrt{s}=13$ TeV using the ATLAS detector}, Phys. Rev. D 94, 032005
\bibitem{ref:monoJet} ATLAS Collaboration, \emph{Search for dark matter and other new phenomena in events with an energetic jet and large missing transverse momentum using the ATLAS detector}, ATLAS-CONF-2017-060.

\bibitem{ref:monoV} ATLAS Collaboration, \emph{Search for dark matter produced in association with a hadronically decaying vector boson in pp collisions at $\sqrt{s}=13$ TeV with the ATLAS detector}, Phys. Lett. B 763 (2016) 251, arXiv:1608.02372 [hep-ex].

\bibitem{ref:dijets} ATLAS Collaboration, \emph{Search for new phenomena in dijet events using 37 fb$^{-1}$ of pp collision data collected at $\sqrt{s}=13$ TeV with the ATLAS detector}, CERN-EP-2017-042, arXiv:1703.09127 [hep-ex].

\bibitem{ref:DMsummary} ATLAS Collaboration, \emph{Summary plots from the ATLAS Exotic physics group}, \url{https://atlas.web.cern.ch/Atlas/GROUPS/PHYSICS/CombinedSummaryPlots/EXOTICS/index.html}

\end{thebibliography}
\end{document}